\documentclass{article}
\usepackage{deluxetable}
\usepackage{graphicx}
\usepackage{ amssymb }

\newcommand{\affil}[1]{$^{\rm #1}$}
\usepackage[authoryear]{natbib}
\bibpunct{ (}{)}{;}{a}{}{,}
\usepackage{graphicx}
\date{} 
\newcommand{\kms}{\mbox{km\,s$^{-1}$}}
\def\kms {\ifmmode{{\rm ~km~s}^{-1}}\else{~km~s$^{-1}$}\fi}
\def\lsun {\ifmmode{{\rm ~L}_\odot}\else{~L$_\odot$}\fi}

\def\sqdeg {\,deg$^2$}

\def\ujybm {\,$\mu$Jy/beam}

\newbox\grsign \setbox\grsign=\hbox{$>$} \newdimen\grdimen \grdimen=\ht\grsign
\newbox\simlessbox \newbox\simgreatbox
\setbox\simgreatbox=\hbox{\raise.5ex\hbox{$>$}\llap
 {\lower.5ex\hbox{$\sim$}}}\ht1=\grdimen\dp1=0pt
\setbox\simlessbox=\hbox{\raise.5ex\hbox{$<$}\llap
 {\lower.5ex\hbox{$\sim$}}}\ht2=\grdimen\dp2=0pt

\def \etal {\rm ~{\it \etal},~}

\def\apj {{\it Ap.~J.}}
\def\apjl {{\it Ap.~J.\ (Letters)}}
\def\apjs {{\it Ap.~J.\ Suppl.}}
\def\aj {{\it A.~J.}}

\def\aap {{\it Astr.~Ap.}}

\def\mnras {{\it MNRAS}}

\def\nat {{\it Nature}}
\def\pasa {{\it PASA}}

\title{\large\bf\flushleft {Discovering the Unexpected in Astronomical Survey Data}}
\author{\parbox{\textwidth}{\flushleft
\vspace{-0.5cm}
{\it
Ray P.\ Norris\affil{1,2}
 \\
{\small \affil{1}\,Western Sydney University, Locked Bag 1797, Penrith South, NSW 1797, Australia}\\
{\small \affil{2}\,CSIRO Astronomy \& Space Science, PO Box 76, Epping, NSW 1710, Australia}\\
\vspace{0.4cm}
}}}
\begin{document}
\twocolumn[
\begin{changemargin}{.8cm}{.5cm}
\begin{minipage}{.9\textwidth}
\vspace{-1cm}
\maketitle
%
%

{\bf Abstract: 
Most major discoveries in astronomy are unplanned, and result from surveying the Universe in a new way, rather than by testing a hypothesis or conducting an investigation with planned outcomes. For example, of the 10 greatest discoveries made by the Hubble Space Telescope, only one was listed in its key science goals.  So a telescope that merely achieves its stated science goals is not achieving its potential scientific productivity. 

Several next-generation astronomical survey telescopes are currently being designed and constructed that will  significantly expand the volume of observational parameter space, and should in principle discover unexpected new phenomena and new types of object. However, the complexity of the telescopes and the large data volumes mean that these discoveries are unlikely to be found by chance. Therefore, it is necessary to  plan explicitly for these unexpected discoveries in the design and construction of the telescope.  Two types of discovery are recognised: unexpected objects, and unexpected phenomena.

This paper argues that next-generation astronomical surveys require an explicit process for detecting the unexpected, and  proposes an implementation of this process. This implementation addresses both types of discovery, and  relies heavily on machine-learning techniques, and also on theory-based simulations that encapsulate our current understanding of the Universe to compare with the data.


}

\medskip{\bf Keywords:} telescopes --- surveys

\medskip
\medskip
\end{minipage}
\end{changemargin}

]

\section{Introduction}
\citet{popper} described the scientific method as a process in which theory is used to make a prediction which is then tested by experiment. 
That model, and its principle of ``falsifiability''.  remains the gold standard of the scientific method, and probably drives the majority of scientific progress. Notable recent successes include the discovery of the Higgs boson \citep{higgs} and the detection of gravitational waves \citep{abbott16}. Conversely,  models such as string theory are sometimes criticised \citep[e.g.][]{woit11} for being unfalsifiable, and thus failing to adhere to this Popperian scientific method. 

However, the Popperian scientific method is not the only one, and a number of other modes of scientific discovery have been proposed, notably by \citet{kuhn}. For example, science may also proceed through a process of ``exploration'' \citep[e.g.][]{harwit81}, in which experiments or observations are carried out in the absence of a compelling theory, in order to guide the development of theory. 

Astronomy has largely developed through a process of exploration. For example, the Hertzsprung-Russell diagram \citep{HR} was an observationally-driven idea of representing data, that led to the development of models of stellar evolution and ultimately nuclear  fusion. In another example, the expanding Universe was discovered when Hubble plotted redshifts of galaxies against their brightness \citep{hubble29}. More recently, the Hubble Deep Fields \citep{williams96,williams00} were primarily motivated by a desire to explore the early Universe, rather than testing specific models or hypotheses.

\subsection{The History of Astronomical Discovery}

Astronomical discovery has often occurred as a result of technical innovation, resulting in the Universe being observed in a way that was not previously possible. Examples include the development of larger telescopes, or the opening up of a new window of the electromagnetic spectrum.  More generally, we may define an n-dimensional  parameter space whose n orthogonal axes correspond to observable quantities (e.g. frequency, sensitivity, polarisation, colour, spatial scale, temporal scale). Some parts of this parameter space have been well-observed and have already yielded their discoveries, whereas some parts of this space have not yet been observed, and new discoveries may lie in those unsampled parts of the parameter space, presumably available to new instruments able to sample that region of the parameter space. Most ``accidental'' or ``serendipitous'' discoveries result  from observing a new part of this parameter space \citep{harwit03}.

We may therefore broadly divide astronomical discoveries into (a) those which were made according to the Popperian model, in which a model or hypothesis is being tested (the known-unknowns), and (b) those which have resulted  from observing the Universe in a new part of the parameter space, resulting in unexpected discoveries (the unknown-unknowns). Of course, an experiment may often be planned to test a hypothesis, but in doing so stumbles across an unexpected discovery. A classic example of this is the discovery of pulsars\citep{hewish68} discussed in \S\ref{pulsar}.

\begin{figure}[h]
\begin{center}
\includegraphics[width=8cm, angle=0]{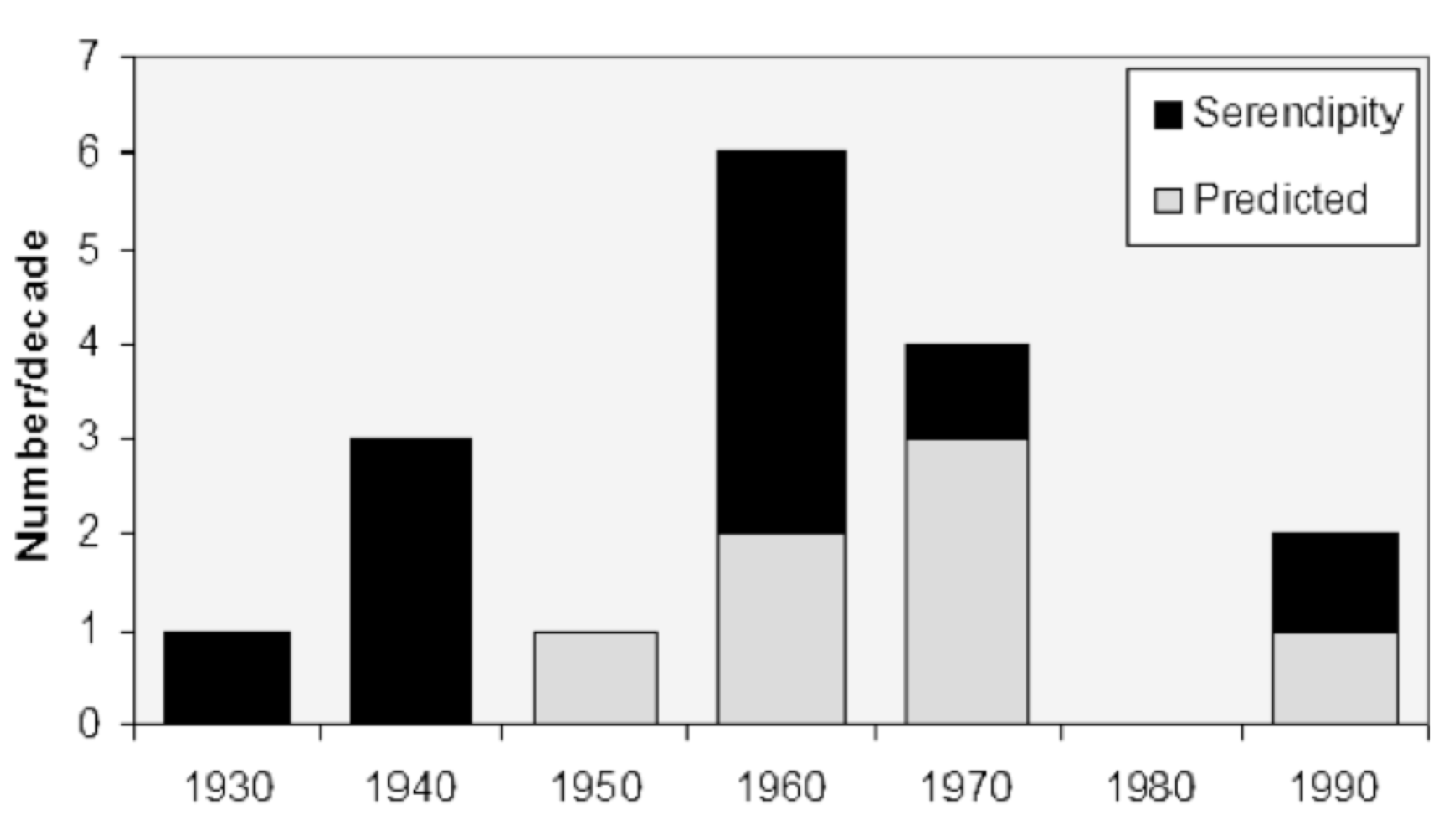}
\caption{A plot of recent major astronomical discoveries, taken from \citep{ekers09}, of which seven were ``known-unknowns'' (i.e. discoveries made by testing a  prediction) and ten were ``unknown-unknowns'' (ie. a serendipitous result found by chance while performing an experiment with different goals). The data in this plot are taken from 
\cite{wilkinson04}. 
}
\label{ekers}
\end{center}
\end{figure}

Several studies \citep{harwit81, wilkinson04, wilkinson07, kellermann09, ekers09, wilkinson15} have shown that 
at least half the major discoveries in astronomy are unexpected, and are typically made by surveying the Universe in a new way, rather than by testing a hypothesis or conducting an investigation with planned outcomes. For example, Figure \ref{ekers} show the result of examining  \citep{ekers09} of
 17 major astronomical discoveries in the last 60 years. Ekers concluded that only seven resulted from systematic observations designed to test a hypothesis or probe the nature of a type of object. The remaining ten were unexpected discoveries resulting either from new technology, or from observing the sky in an innovative way, exploring uncharted parameter space.  In particular, experience has shown that unexpected discoveries often result when the sky is observed to a significantly greater sensitivity, or a significantly new volume of observational parameter space is explored.
 
\subsection{This paper}
In Section 2 of this paper I discuss the opportunities and challenges to making unexpected discoveries in the high data volumes and high complexity of next-generation astronomical surveys, and argue that surveys need to plan explicitly for these discoveries if they are to be successful. Section 3 proposes a process for discovering unexpected objects in astronomical surveys,  and Section 4 proposes a process for discovering unexpected phenomena in astronomical surveys. Section 5  describes some preliminary attempts to implement and test some of these approaches and suggests some future directions. 

To focus the discussion, this paper uses the ``Evolutionary Map of the Universe'' survey \citep[EMU:][]{norris11} as an exemplar of next-generation surveys, but the broad conclusions and process will be relevant to all next-generation astronomical surveys.

\section{The process of astronomical discovery}
Astronomy is currently enjoying a boom in new surveys, with several next-generation astronomical survey telescopes planned, which will undoubtedly open up large new swathes of observational parameter space, potentially resulting in a large number of unexpected discoveries. 

There are two quite different types of unexpected discovery:
\begin{itemize} 
\item Type 1: Discoveries of new types of object (e.g. pulsars, quasars), identified as anomalies or unexpected objects in  images or catalogs;
\item Type 2: Discoveries of new phenomena (e.g. HR diagram, the expanding Universe, dark energy), identified as anomalies in the distributions of properties of objects. These are identified when the results of experiment are compared to theory (or perhaps to other observations) in some suitable parameter space. 
\end{itemize}

\subsection{Case study 1: The Hubble Space Telescope}
\label{HST}

\begin{table*}
\caption{Major discoveries made by the Hubble Space Telescope ({\it HST}). Of the {\it HST}'s ``top ten'' discoveries (as ranked by National Geographic magazine), only one was a key project used in the {\it HST} funding proposal \citep{Lallo}.  A further four projects were planned in advance by individual scientists but not listed as key projects in the {\it HST} proposal. Half the ``top ten'' {\it HST} discoveries were unplanned, including two of the three most cited discoveries, and including the only {\it HST} discovery (Dark Energy) to win a Nobel prize. This Table was previously published by \cite{norris15}.}
\vspace{2mm}
\begin{tabular}{llllll}
\hline
Project  & Key  & Planned? & Nat Geo 
 & Highly  & Nobel \\
  & Project? & &top ten? & cited? & Prize? \\
\hline
Use cepheids to improve value of $H_{0} $&\checkmark  & \checkmark &\checkmark  & \checkmark & \\
UV spectroscopy of ig medium  &\checkmark  & \checkmark & & & \\
Medium-deep survey &\checkmark  & \checkmark & & & \\
Image quasar host galaxies & &\checkmark  & \checkmark & & \\
Measure SMBH masses & & \checkmark &\checkmark  & & \\
Exoplanet atmospheres & & \checkmark &\checkmark  & & \\
Planetary Nebulae & &\checkmark  &\checkmark  & & \\
Discover Dark Energy & & & \checkmark &\checkmark  &\checkmark  \\
Comet Shoemaker-Levy & & & \checkmark & & \\
Deep fields (HDF, HDFS, GOODS, FF, etc) & & & \checkmark &\checkmark  & \\
Proplyds in Orion  & & & \checkmark & & \\ 
GRB Hosts & & &\checkmark  & & \\

\hline
\end{tabular}
\label{HST}

\end{table*}

The science goals that drove the funding, construction, and launch of the Hubble Space Telescope (HST) are listed in the {\it HST} funding proposal \citep{Lallo}.  A further four projects were planned in advance by individual scientists but not listed as key projects in the {\it HST} proposal. Conveniently, the National Geographic magazine selected the ten major discoveries of the HST  \citep{natgeo}, resulting in an admittedly subjective  "top ten" list of HST discoveries. So we may compare the actual achievements of the HST against its planned achievements. Of these ten greatest discoveries by HST, only one was listed in its key science goals.   In particular, the unplanned discoveries  include two of the three most cited discoveries, and the only {\it HST} discovery (Dark Energy) to win a Nobel prize.

This example suggests that science goals are   poor predictors of the discoveries to be made with a new telescope, and if a major new telescope merely achieves its stated science goals, it is probably performing well below its potential scientific productivity. \cite{wilkinson04} express this  idea succinctly as {\it What a radio telescope was built for is almost never what it is known for.}

\subsection{Case Study 2: the Discovery of Pulsars}
\label{pulsar}
The Nobel-prize-winning discovery of pulsars by Jocelyn Bell occurred when a talented and persistent PhD student observed the radio sky for the first time with high time resolution, to study interstellar scintillation.  By observing at high time resolution,  she expanded the observational parameter space. She also knew her instrument intimately, enabling her to recognise that ``bits of scruff'' on the chart recorder could not be due to terrestrial interference, but represented a new type of astronomical object. As a result, she discovered pulsars. She describes the process in detail in \citet{bell09}.

The following critical elements were essential for this discovery.
\begin{itemize}
\item she explored a new area of observational parameter space
\item she knew the instrument well enough to distinguish interference from signal
\item she examined all the data by eye
\item she was observant enough to recognise something unexpected
\item she was open minded, and prepared for discovery
\item she was within a supportive environment (i.e. one that was accustomed to making new discoveries).
\item she was persistent 
\end{itemize}

The value of the last three items should not be underestimated. When a PhD student obtains an observational result that differs from previous results or from conventional wisdom, there is a strong temptation to ascribe the difference to an error in the data.

\subsection{Case Study 3: The Evolutionary Map of the Universe}

\begin{figure}[h]
\begin{center}
\includegraphics[width=8cm, angle=0]{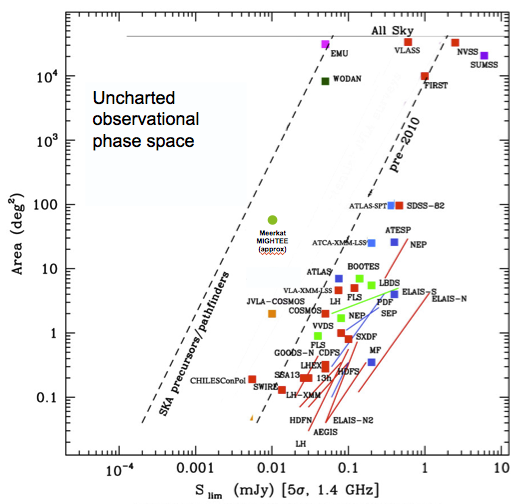}
\caption{Comparison of existing and planned deep 20 cm radio continuum surveys, adapted from a diagram originally drawn by Isabella Prandoni. The horizontal axis shows the  5-$\sigma$ sensitivity, and the vertical axis shows the sky coverage. The right-hand diagonal dashed line shows the approximate envelope of existing surveys, which is largely determined by the availability of telescope time.  Surveys not at 20cm are represented at the equivalent 20 cm flux density, assuming a spectral index of -0.8.
The squares in the top-left represent the new radio surveys discussed in this paper.  The Square Kilometre Array \citep{dewdney09} will  hopefully  conduct even larger surveys in the next decade, extending well to the left of EMU, but such plans are not yet concrete.
}
\label{surveys}
\end{center}
\end{figure}

 Figure 1 shows the main radio surveys, both existing and planned, at frequencies close to 1.4 GHz. The largest existing radio survey, shown in the top right, is the wide but shallow NRAO VLA Sky Survey \citep[NVSS:][]{condon98}. The most sensitive existing radio survey is the deep but narrow JVLA-SWIRE (Lockman hole) observation in the lower left \citep{condon12}. 
Existing surveys are bounded by a diagonal line that roughly marks the limit of available  time on current-generation radio telescopes.  

Many discoveries have been triggered by those surveys shown in Figure 1, ranging from the rare but paradigm-shifting discoveries (e.g. the radio-far-infrared correlation \citep{vdk71} ) to the  numerous  minor but still significant discoveries (e.g. the Infrared-Faint Radio Sources \citep{norris06}, which are now known to be very-high-redshift radio galaxies \citep{garn08, herzog14, collier14}). In the absence of any evidence to the contrary, Occam's razor would suggest that this diagram is uniformly populated with significant discoveries. Therefore, the unexplored region of observational parameter space to the left of the line presumably contains as many  potential new discoveries per unit parameter-space as the region to the right. Radio surveys of that region should therefore yield many important discoveries, provided they are equipped to do so.

Within that unexplored region of parameter space are several planned next-generation radio surveys, the largest of which, in terms of numbers of sources detected, is  EMU \citep[Evolutionary Map of the Universe;][]{norris11} which will use the Australian SKA Pathfinder \citep{johnston08}, to survey 75\% of the sky to a sensitivity of 10\,\ujybm\ rms.
Only a total of about 10\,\sqdeg\  of the sky has been surveyed at 1.4\,GHz to this sensitivity, in fields such as the {\it Hubble}, ATLAS, and COSMOS fields. 
EMU is the largest  radio continuum survey so far, and will detect about 70 million galaxies, compared to the 2.5 million detected over the entire history of radioastronomy. 
Not only will EMU have greater sensitivity than  previous large-area surveys, but it will also have better resolution, better sensitivity to extended emission, and will
measure spectral index and, courtesy of the POSSUM project \citep{gaensler10}, polarisation for the strongest 
sources. 

EMU will therefore significantly expand the volume of observational parameter space, so  in principle should discover unexpected new phenomena and new types of object. 

However, the complexity of ASKAP and the large data volumes mean that it may be non-trivial to identify them. 
For example, in the list above of  critical elements which led to the discovery of pulsars, EMU can satisfy all those elements except (a) knowing  the instrument well enough to distinguish interference or artefacts from signal, (b)
being able to examine all the data by eye, and (c) being able to recognise something unexpected

For (a), it is likely  that no  human  will be sufficiently familiar with ASKAP  to distinguish subtle astrophysical effects from subtle instrumental artefacts. Any process to detect unexpected astrophysical effects is likely to detect unexpected artefacts. Rather than expecting to identify these a priori, it is likely that we will have to learn to identify them in the data, and then trace their source {\it a posteriori}. This process is likely to be an important component of the process of discovering the unexpected.

For (b), the petabyte data volumes from ASKAP mean that it will be  impossible for an astronomer to sift through the data, looking for something unusual. Instead, the only way of extracting science from large volumes of data is to interrogate the data with a well-posed question, such as `plot the specific cosmic star formation rate of star-forming galaxies as a function of redshift'. So there is a danger that projects like EMU will produce good science in response to such well-posed questions (the "known-unknowns"), and thus achieve their science goals, but will miss the 90\% of discoveries that are unexpected  (the "unknown-unknowns").

The final element (c), of being able to recognise something unexpected, is perhaps the hardest element. While the human brain has been exquisitely tuned  by millions of years of evolution to notice anything unexpected and potentially dangerous, if we can't sift through the data by eye, then we must rely on tools to detect the unexpected, and such tools do not currently exist.


 On the other hand, if we donÕt make the unexpected discoveries, then we will probably miss out on the most important science results from these telescopes. 
We have therefore started a  project within EMU (named Widefield ouTlier Finder, or WTF) to develop techniques for mining large volumes of astronomical data for the unexpected, using  machine-learning techniques and algorithms.

 \subsection{The value of Science Goals}
\label{goals}
New telescopes or surveys are usually justified by their science goals. For example, the EMU project \citep{norris11} is justified by 16 key science projects with goals such as measuring the star formation rate density over cosmic time, studying AGN evolution and the role of AGN feedback, and
making independent measurements of fundamental cosmological parameters. However, as demonstrated above in the case of the HST, the major discoveries made with a new telescope or survey are not usually represented by such science goals. 

However, science goals are still important for two reasons. First, they represent use cases. If a telescope is built that is able to address challenging science goals, then it is likely to be a high-performing telescope. Second, much of astronomy advances not by spectacular major discoveries, but by the  incremental science that is usually encapsulated in science goals. Such incremental advance is also very important, and, unlike serendipitous discoveries, represents a predictable outcome from a new telescope.

For example, EMU will hopefully advance the knowledge of galaxy evolution by measuring the evolution of the cosmic star formation rate, the evolution of active galactic nuclei, and the feedback processes that link them, and this will no doubt result in many worthwhile and highly-cited papers. However, these may be dwarfed in impact by the unexpected discoveries.

\section{Type 1 Discoveries: unexpected objects}

EMU is expected to detect about 70 million objects, compared
to the current total of $\sim$ 2.5 million known radio sources. Since the 70 million objects will probably include new unexpected classes of radio source, it is  important for EMU to plan to identify new classes or phenomena, rather than hoping to stumble across them. EMU will do so through its Widefield ouTlier Finder (WTF) project, which has the explicit goal of discovering the unexpected.

\begin{figure}[h]
\includegraphics[width=8cm, angle=0]{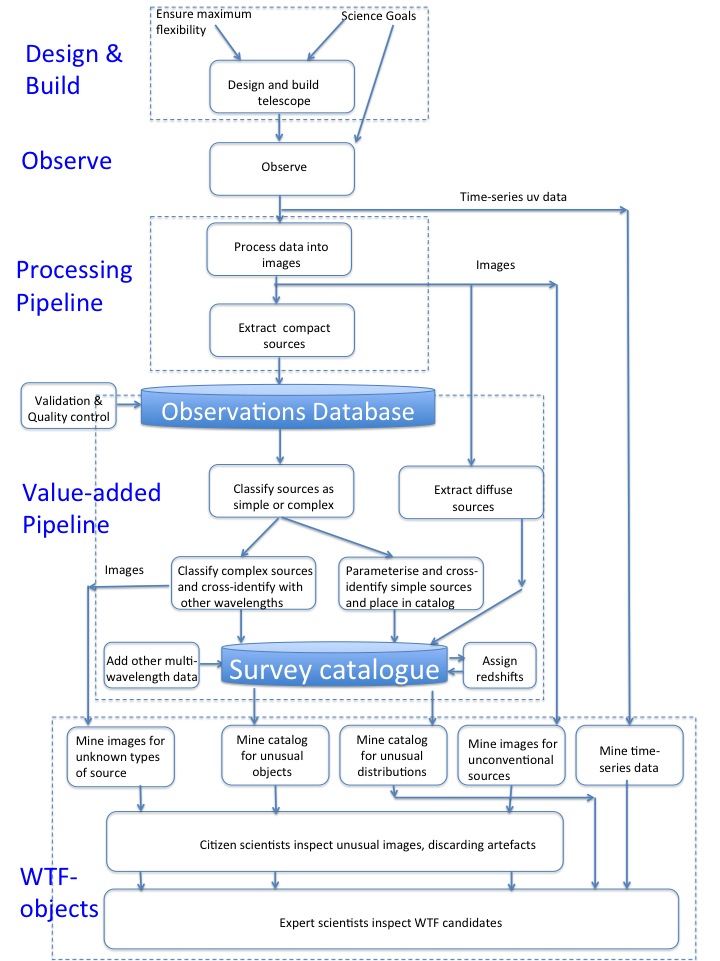}
\caption{The flowchart for discovering unexpected objects in EMU.
}
\label{flowchart1}
\end{figure}

This section describes how the WTF project will make Type 1 discoveries (unexpected objects),  An overview is shown in Figure \ref{flowchart1}  and the following subsections address each of the steps in that flowchart. Although this is designed for EMU,  the broad approach is applicable to any survey.

\subsection{Design and Construction}
As discussed in \S\ref{goals}, 
the construction of any new telescope must necessarily be designed to optimise its performance for specific science goals. However, it is important not to design and  build it so it can {\it only} achieve those goals, because that would limit its ability to discover the unexpected. Instead, it is important to maximise flexibility.  The design of the telescope therefore needs  to maximise the ultimate scientific productivity, in addition to achieving the specific science goals.

Similarly, it is sometimes necessary to process the data to reduce the volume of data to that which is necessary to achieve the science goals, discarding the excess. For example, ASKAP will generate about 70 PB of calibrated correlated time-series data each year, which is then processed into images occupying only about 4 PB per year. It is not economically possible to store all the time-series spectral-line data, and so that data is discarded.

Discarding  excess data is sensible if all the information is present in the images. However, processing the time-series data to produce the images is a lossy process, and the discarded  information may well be the key to an unexpected discovery. So reducing the data volume by keeping only processed data should be avoided as much as possible.

Even when time-series data must be discarded, it can still be searched in real time for time-varying phenomena such as fast radio bursts\cite{lorimer07}. In the case of EMU, this search is undertaken by partner projects CRAFT \citep{macquart10} and VAST \citep{murphy13}.

\subsection{Observations}

Discoveries are thinly distributed through the observational parameter space.  We cannot predict where they lie, and it is difficult to quantify the volume of parameter space being explored, but the probability of making an unexpected discovery is presumably proportional to the volume of new parameter space being explored. 
The observations should therefore be optimised, not only for the specific science goals, but also to maximise the volume of new observational parameter  space being explored, which means maximising the sensitivity to
 poorly explored parameters such as circular polarisation, time variability,  diffuse emission, etc.

\subsection{Data Processing and Compact Source Extraction}
The first stage of ASKAP data processing, performed by  the ASKAPSOFT suite of software, is to calibrate the time-series data, Fourier transform it into image data, and then deconvolve it. The resulting images are then placed in the observations database (called CASDA) for storage and retrieval by users.

 It is important that this process makes as few assumptions as possible about the nature of the objects being detected. For example, we know that the vast majority of objects detected by EMU will be less than one arcmin in extent, and so it is tempting to discard the shortest baselines corresponding to spatial scales larger than this. However, to do so will be to guarantee that EMU will not detect any objects larger than this scale, thereby limiting the volume of observation parameter space being explored.

The ASKAPSOFT real-time processing pipeline includes source extraction software to identify and measure the parameters of compact sources in the radio images. The algorithm for doing so is still being refined and tested against other source finders \citep{hopkins15}, but is optimised for sources that are unresolved or less than  a few beamwidths in extent. The software will measure the extent of each component (an ``island'') and fit gaussians to the peaks within the island. The measured parameters from this process are stored in a  table in CASDA for storage and retrieval by users.

Diffuse sources will not normally be discovered by this process, but will be extracted in offline processing (see \S \ref{diffuse}).


\subsection{Data Validation}
 
The first stage of EMU data validation takes place in near-real-time to flag data which are affected by radio-frequency interference or hardware malfunctions. A second stage of validation is conducted on each set of observations by the EMU science survey team, checking for image artefacts, calibration errors, etc. It is important to ensure that this process does not also reject data containing unexpected discoveries. For example, a strong radio burst might be misinterpreted as interference. However, an astrophysical radio burst will take place in the far field of ASKAP, while interference generally takes place in the near field. Interference can therefore be distinguished from radio bursts by testing whether the parameters on different baselines are consistent with an astrophysical source. It is therefore important that data validation techniques use such  sophisticated tests rather than simple amplitude threshold tests.

\subsection{Diffuse Source Extraction}
\label{diffuse}
 The source extraction algorithm in ASKAPSOFT is not expected  to detect diffuse emission, such as cluster haloes and supernova remnants, which are notoriously difficult to detect automatically. A number of algorithms \citep[e.g.][]{dabbech15, butler16, riggi16} are under development for automatically detecting diffuse sources in radio-astronomical images.

\subsection{Classification of Sources as Simple or Complex}

About 90\% of EMU sources will consist of a single radio component with no nearby radio component with which it might be associated. I term these `simple' sources. Physically, these are likely to be star-forming galaxies, low-luminosity AGN, or young radio-loud galaxies typically classified as Gigahertz-peaked spectrum (GPS) or compact steep spectrum (CSS). The first stage of classification and identification is to identify such sources from their radio morphology alone.  This separation into simple and complex sources will be achieved in EMU using a machine-learning algorithm, currently under development  \citep{park17}. It is likely that the final algorithm will use one of Logistic Regression, a Support Vector Machine,
or a Neural Network binary classification. 

 The resulting simple sources will then be matched to optical/infrared catalogues using a likelihood ratio (LR) technique 
 \citep{sutherland92,weston17}.

The remaining sources, which we term `complex', must be classified and cross-identified in a more sophisticated process. 

\subsection{Source classification and cross-identification of complex sources}
Classifying the morphology of radio sources, and cross-identifying them with their counterparts at optical/infrared wavelengths, might be regarded as being two separate processes. However, two nearby unresolved radio components might either be the two lobes of an FRII radio source, or the radio emission from two unassociated star-forming galaxies. Only by cross-identifying with multiwavelength data, particularly optical/infrared data, can these two cases be distinguished, since the pair of star forming galaxies will have an infrared host galaxy coincident with each of the radio components, whereas the host of the FRII is likely to lie between them.

Whilst this process is easy for the expert human, the 7 million complex sources expected to be detected by EMU pose a significant challenge. Several techniques are being evaluated, using the $\sim$5000 sources in the ATLAS data set \citep{norris06, middelberg08, hales14, franzen15} as a testbed, as follows:

\begin{itemize}
\item All sources are cross-identified and classified by eye, to provide a training and validation set
\item The sources are being cross-matched by citizen scientists in the Radio Galaxy Zoo project \citep{banfield16}
\item A Bayesian approach is being developed \citep{fan15}
\item A variety of machine-learning approaches are being explored, both supervised and unsupervised.
\end{itemize}

\subsection{The Survey Catalogue}
After cross-matching and classification, all sources detected in the survey are placed in the survey catalogue, which for EMU is called the EMU Value-Added Catalogue (EVACAT ).  To each source are added other available data such as redshifts and other multiwavelength data. Many of the redshifts are not spectroscopic, but are photometric redshifts or ``statistical redshifts''\citep{norris11} which are best expressed as a probability distribution function rather than as a single value.

\subsection{Mining images for unexpected objects}
 The source extraction algorithm in ASKAPSOFT is not expected  to detect unconventional sources. An example of an unconventional source might be a ring of emission several arcmin in diameter but with an amplitude of only half the rms noise level in any one pixel. Such a structure would be invisible in the image to the human eye, or to a conventional source extraction code, but would be easily detectable at a high level of significance using a  suitable matched filter, such as a Hough transform \citep{hollitt12}. Many other examples of potential diffuse and unconventional sources may be imagined.

To detect such sources, the WTF pipeline will retrieve images from CASDA and apply a number of different algorithms in parallel. Detecting sources with unconventional morphology is much harder and is the subject of continuing research, and several algorithms such as self-organised maps \citep{geach12} are currently being explored.

\subsection{Mining the catalogue for unexpected objects}

The catalogue will be searched  for properties of objects in an n-dimensional plot with axes such as flux density, spectral index, and IR-to-radio ratio.  Known types of object (e.g. stars, galaxies, quasars) will appear as clusters in this parameter space. Algorithms are being explored that will search the parameter space for clusters of objects that do not correspond to known types of objects.  
Although targeted specifically at EMU, such approaches are expected to have broad applicability to astronomical survey data.
 
\section{Type 2 Discoveries: Unexpected Phenomena}
Some unexpected discoveries are made when the properties of a sample of objects differ from those predicted by theory in some unexpected way. For example, dark energy was discovered \citep{riess98, perlmutter99} when the relation ship between the brightness and redshift of type 1A supernovae  failed to follow the expected distribution predicted by theory. Here I describe an approach in which the data is tested against theory. Although it resembles the standard Popperian technique, it differs in that what is being tested is the sum of our understanding of the Universe, rather than any particular theory.

%

A common way of testing theories is to derive some physically meaningful quantity, such as a luminosity function, and then compare that with the theoretical luminosity function predicted by theory. Such an approach has the advantage of yielding results which are easily compatible with other observations and other theories. It has the disadvantage that observational data has to be corrected for incompleteness, and this is often difficult to do accurately. For example, to calculate the radio luminosity function of radio sources, and compare it with other derived radio luminosity functions, \cite{mao12} needed to correct the data not only for a variable radio sensitivity across the field, but for the incompleteness of the optical spectroscopy survey that produced the necessary redshifts. It is very difficult to account for all the selection effects accurately.

These various sources of incompleteness, which I label the ``window function'', are generally well-understood and well-determined. For example, \cite{mao12} were able to use a map of the sensitivity across the radio image, and a plot of the sensitivity of the redshift survey as a function of magnitude. Thus for a hypothetical source of a given optical magnitude and position, it is trivial to calculate the probability of it appearing in the catalogues with a  measured redshift. The converse process is much harder - correcting the catalogue for these effects requires a number of approximations.
  It is likely that the differences between different measurements of this radio luminosity function \citep[e.g.][]{mao12, mauch07, padovani11} is primarily caused by these approximations.

An alternative  to correcting the data is to compare it with physically realistic models,  is to use the theory to simulate the observations, and then apply the window function to result in simulated data that can be compared with the original data. Of course, a particular simulated galaxy will not coincide with a  particular real galaxy, and so it is necessary to compare the statistical properties of the simulate data to those of the real data. But this comparison can be done in a parameter space which is close to that of of the real data (e.g. source counts as a function of flux density in the survey volume), rather than transforming it to a physically meaningful parameter space (e.g source counts as a function of luminosity in an idealised volume). This may be regarded as a Bayesian process, in that the theory is being used to predict the data, rather than the theory being inferred from the data.

In the case of searching for the unexpected, the simulations are being used to encapsulate our current understanding of astrophysics so that they can be compared with the data, to see if the data is consistent with our current understanding. Any significant difference between the two either represents an error in the data or simulation, or an unexpected discovery.

\begin{figure}
\begin{center}
\includegraphics[width=8cm, angle=0]{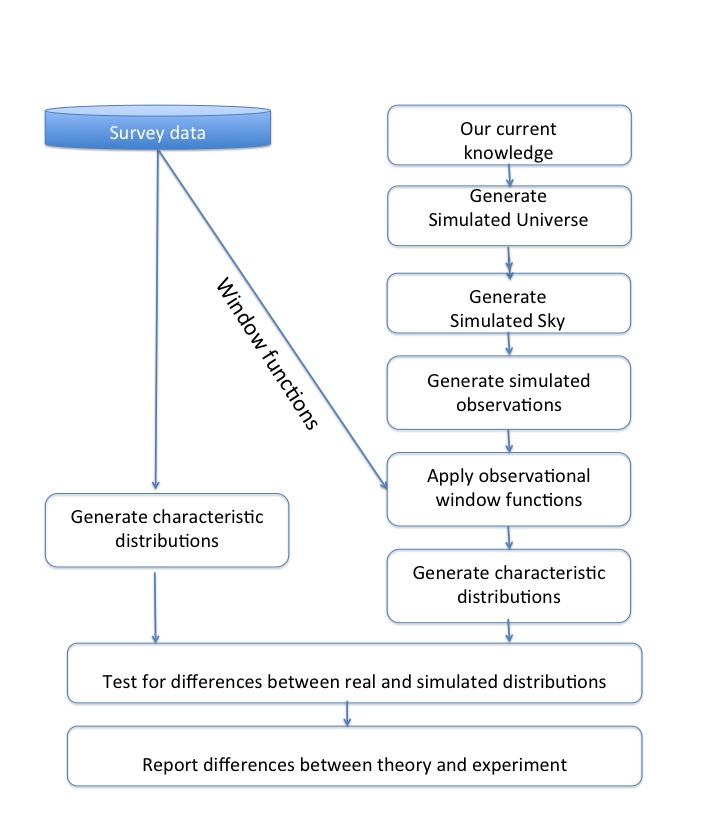}
\caption{The flowchart for discovering unexpected phenomena in the EMU WTF project
}
\label{flowchart2}
\end{center}
\end{figure}

This process is shown in Figure \ref{flowchart2}, and includes the following steps. 
The starting point is a simulation, such as the Millennium Simulation \citep{springel05} which encapsulates our knowledge about cosmology and galaxy formation. From this is generated a simulated sky, using our knowledge of the observed properties of galaxies. Tools such as the Theoretical Astrophysical Observatory \citep[TAO:][]{bernyk16} are designed to do this. However, TAO does not yet generate a radio sky, and so a simulated radio sky must be generated from the TAO sky using a  semi-empirical model of radio sources. 
The model sky is them converted to a simulated observed sky using observational constraints such as sensitivity and resolution. The window function is then applied including factors such as area of sky observed, and any varying sensitivity across the observations.

A characteristic distribution is a representation of the observational or simulated data  which represents the data in some particular parameter space. Well-known examples include source count plots and angular power spectrum, but in principle almost any observational quantity can be plotted against any other, and there is no need for these plots to be confined to two dimensions. To systematically search for unexpected deviations of theory from data, all combinations of observational quantities need to be searched by algorithms which will report significant anomalies to the user.

\begin{figure}
\begin{center}
\includegraphics[width=8cm, angle=0]{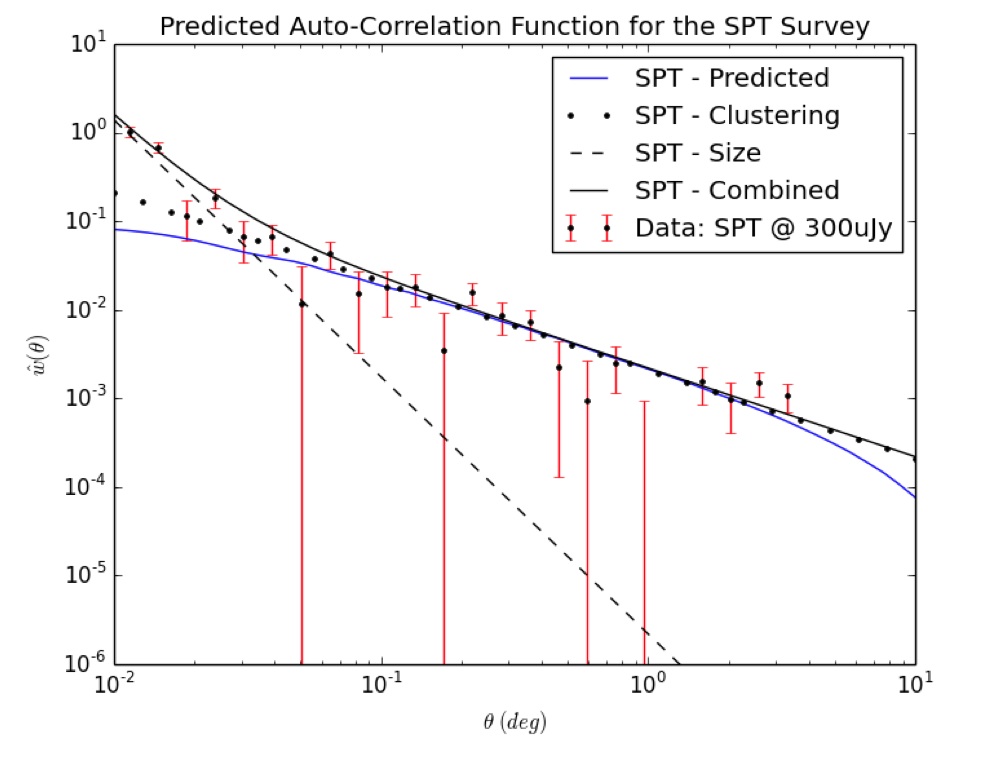}
\caption{The angular power spectrum for radio sources in the SPT field, taken from \cite{rees17}. Points with error bars are the measured angular power spectrum of the data obtained by \cite{obrien}, and the blue line shows the distribution predicted by the semi-empirical model described in the text. The dotted line shows the cosmological signal predicted by $\Lambda$CDM,and the dashed line show the effect of radio source size and double radio sources. The solid black line is the sum of these latter two predictions.}
\label{rees}
\end{center}
\end{figure}

A simple example of this process, taken from \citep{rees17} is shown in Figure \ref{rees}. Here, the characteristic distribution is the angular power spectrum for radio sources in the SPT (South Pole Telescope)  field, using the radio observations described by \cite{obrien}. The simulated data were based on the Millennium Simulation, from which a simulated sky of galaxies was generated using the TAO  tool. From this, a radio sky was generated as described by \cite{rees17} using semi-empirical assumptions about the properties of radio sources based on the zFOURGE survey \citep{rees16}. In this case, the observational data were corrected for the window function, but the correction could equally well be applied to the simulation data. In this case, the data are found to be consistent with the simulation.

\section {Preliminary Attempts, and Future Directions}

To test the ideas driving this paper, a data challenge was constructed on the Amazon Web Services (AWS)  cloud platform \citep{crawford16}.
 Initially, we wanted to see which algorithms and techniques are best at finding unexpected results, and so we constructed a number of data challenges in which data sets (both real and simulated, and both images and tabular data) are constructed with simulated unexpected discoveries (known as ``eggs'') buried in them. We  then invited machine learning groups to try out their algorithms to see if they could find the simulate eggs. 

This approach was less successful than expected, for the following reasons: 
\begin{itemize}
\item We had underestimated the difficulty of non-astronomers engaging in this project. Specific difficulties included file formats, and the need to present the problem in a way accessible to non-astronomers
\item Lack of personpower: such a project requires dedicated resources
\item The most important factor was that discovering the unexpected is harder than expected.
\end{itemize}
As a result of that experiment, it was clear that a more systematic approach was needed, resulting in the process described in this paper. By breaking the problem down into building blocks, it also makes it a more tractable problem for a team-based approach. Furthermore, many of the building blocks are important tools in their own right that are necessary to extract even the known-unknowns from EMU (e.g. classification and cross-identification of radio sources). 

Other avenues of research are also likely. For example, it is likely that in the Search for Extra-terrestrial Intelligence (SETI), any detected civilisation is likely to be so much more advanced than ours \citep{norris99} than we might not recognise an intelligent signal. A better strategy may be simply to look for signals that are different from those that we expect from known astrophysical processes. In that case, a search for SETI reduces to searching for the unexpected, and can use the process proposed here.

\section{Conclusion}

\begin{itemize}
\item Most major discoveries in astronomy are unexpected.
\item In the past, unexpected discoveries were made serendipitously by users pursuing other goals or exploring the parameter space. However, the complexity of next-generation instruments, and the large volumes of data generated, make it unlikely that they will make such unexpected discoveries. Instead, telescopes must be designed explicitly  to maximise their ability to discover the (potentially more important){ \bf unknown} science goals. 
\item The use of science goals when planning a new telescope are valuable as ``use cases'' for helping design a good project, and are also likely to provide much of the incremental science that results from a successful project, but they are unlikely to represent the most significant science output from the telescope.
\item With the exception of telescopes designed specifically to answer a particular science question, telescopes that merely achieve their stated science goals have probably failed to capture the most important scientific discoveries available to them
\item Because of the complexity and large data volumes of next-generation scientific projects, unexpected discoveries are less likely to happen by chance, but will require software designed to mine the data for unexpected discoveries.
\item Unexpected discoveries may be either Type 1 (unexpected objects) or Type 2 (unexpected phenomena), and it is necessary to design processes to deal with both types.
\item A process has been proposed for finding each of these types in radio survey data, and it is expected that this process may be broadly applicable to other types of astronomical survey.
\end{itemize}

\section*{Acknowledgments}
I thank Laurence Park, Evan Crawford, and Kai Polsterer for valuable discussions. I thank Amazon Web Services for grant EDU\_R\_FY2015\_Q3\_SKA\_Norris that enabled an early prototype to be constructed on  the AWS cloud platform. I also thank ... for helpful comments on an early draft of this paper. I acknowledge the Wajarri Yamatji people as the traditional owners of the ASKAP Observatory site.


\onecolumn

%
%
%


\begin{thebibliography}{}
\bibitem[Abbott et al.(2016)]{abbott16}Abbott, B.P.,  et al. 2016, Phys. Rev. Lett. 116, 061102 
\bibitem[Banfield et al.(2016)]{banfield16} Banfield, J.~K., Andernach, H., Kapi{\'n}ska, A.~D., et al.\ 2016, \mnras, 460, 2376 
\bibitem[Bell-Burnell (2009)]{bell09} Bell-Burnell, J.\ 2009, in 'Accelerating the Rate of  Astronomical DiscoveryÕ, http://pos.sissa.it/cgi-bin/reader/conf.cgi?confid=99 
\bibitem[Bernyk et al.(2016)]{bernyk16} Bernyk, M., Croton, D.~J., Tonini, C., et al.\ 2016, \apjs, 223, 9 
\bibitem[Butler-Yeoman et al.(2016)]{butler16} Butler-Yeoman, T., Frean, M., Hollitt, C.~P., Hogg, D.~W., \& Johnston-Hollitt, M.\ 2016, arXiv:1601.00266 
\bibitem[Collier et al.(2014)]{collier14} Collier, J.~D., Banfield, J.~K., Norris, R.~P., et al.\ 2014, \mnras, 439, 545 
\bibitem[Condon et al.(1998)]{condon98}Condon, J.~J., Cotton, W.~D., Greisen, E.~W., et al.\ 1998, {\it The NRAO VLA Sky Survey} \aj, 115, 1693 
\bibitem[Condon et al.(2012)]{condon12}Condon, J.~J., Cotton, W.~D., Fomalont, E.~B., et al.\ 2012, {\it Resolving the Radio Source Background: Deeper Understanding through Confusion} \apj, 758, 23 
\bibitem[Crawford et al.(2016)]{crawford16} Crawford, E., Norris, R.~P., \& Polsterer, K.\ 2016, arXiv:1611.02829 
\bibitem[Dabbech et al.(2015)]{dabbech15} Dabbech, A., Ferrari, C., Mary, D., et al.\ 2015, \aap, 576, A7 
\bibitem[Dewdney et al.(2009)]{dewdney09} Dewdney, P.~E., Hall, P.~J., Schilizzi, R.~T., \& Lazio, T.~J.~L.~W.\ 2009, IEEE Proceedings, 97, 1482 
\bibitem[Ekers(2009)]{ekers09} Ekers, R.~D.\ 2009, in 'Accelerating the Rate of  Astronomical DiscoveryÕ, http://pos.sissa.it/cgi-bin/reader/conf.cgi?confid=99 
\bibitem[Fan et al.(2015)]{fan15} Fan, D., Budav{\'a}ri, T., Norris, R.~P., \& Hopkins, A.~M.\ 2015, \mnras, 451, 1299 
\bibitem[Franzen et al.(2015)]{franzen15} Franzen, T.~M.~O., Banfield, J.~K., Hales, C.~A., et al.\ 2015, \mnras, 453, 4020 
\bibitem[Gaensler et al.(2010)]{gaensler10} Gaensler, B.~M., Landecker, T.~L., Taylor, A.~R., \& POSSUM Collaboration 2010,  Bulletin of the American Astronomical Society, 42, \#470.13 
\bibitem[Garn \& Alexander(2008)]{garn08} Garn, T., \& Alexander, P.\ 2008, \mnras, 391, 1000
\bibitem[Geach(2012)]{geach12} Geach, J.~E.\ 2012, \mnras, 419, 2633  
\bibitem[Hales et al.(2014)]{hales14} Hales, C.~A., Norris, R.~P., Gaensler, B.~M., et al.\ 2014, {\it ATLAS 1.4 GHz data release 2 - II. Properties of the faint polarized sky} \mnras, 440, 3113 
\bibitem[Harwit(1981)]{harwit81}Harwit, M., 1981, ÒCosmic DiscoveryÓ 
\bibitem[Harwit(2003)]{harwit03}Harwit, M.\ 2003, Physics Today, 56, 38 
\bibitem[ATLAS(2012)]{higgs}ATLAS Collaboration, 2012, Phys.Lett. B716, 1-29
\bibitem[Hertzsprung(1908)]{HR}Hertzprung, E., 1908, Astronomische Nachrichten. 179 (24): 373Ð380. 
\bibitem[Herzog et al.(2014)]{herzog14} Herzog, A., Middelberg, E., Norris, R.~P., et al.\ 2014, \aap, 567, A104 
\bibitem[Hewish et al.(1968)]{hewish68} Hewish, A., Bell, S.~J., Pilkington, J.~D.~H., Scott, P.~F., \& Collins, R.~A.\ 1968, \nat, 217, 709 
\bibitem[Hollitt \& Johnston-Hollitt(2012)]{hollitt12} Hollitt, C., \& Johnston-Hollitt, M.\ 2012, \pasa, 29, 309 
\bibitem[Hopkins et al.(2015)]{hopkins15} Hopkins, A.~M., Whiting, M.~T., Seymour, N., et al.\ 2015, \pasa, 32, e037 
\bibitem[Hubble(1929)]{hubble29} Hubble, E.\ 1929, Proceedings of the National Academy of Science, 15, 168 
\bibitem[Johnston et al.(2008)]{johnston08} Johnston, S., Taylor, R., Bailes, M., et al.\ 2008, Experimental Astronomy, 22, 151 
\bibitem[Kellermann(2009)]{kellermann09}Kellermann, K.I., et al., 2009, in 'Accelerating the Rate of Astronomical http://pos.sissa.it/cgi-bin/reader/conf.cgi?confid=99 
\bibitem[Kuhn(1962)]{kuhn}Kuhn, T. S. 1962, The Structure of Scientific Revolutions. The University of Chicago Press. 
\bibitem[Lallo(2012)]{Lallo}Lallo, M.~D.\ 2012, {\it Experience with the Hubble Space Telescope: 20 years of an archetype} Optical Engineering, 51, 011011 (arXiv:1203.0002) 
\bibitem[Lorimer et al.(2007)]{lorimer07} Lorimer, D.~R., Bailes, M., McLaughlin, M.~A., Narkevic, D.~J., \& Crawford, F.\ 2007, Science, 318, 777 
\bibitem[Macquart et al.(2010)]{macquart10} Macquart, J.-P., et al., \pasa, 27,272. 
\bibitem[Mao et al.(2012)]{mao12} Mao, M.~Y., Sharp, R., Norris, R.~P., et al.\ 2012, \mnras, 426, 3334 
\bibitem[Mauch \& Sadler(2007)]{mauch07} Mauch T., \& Sadler E. M., 2007, \mnras, 375, 931
\bibitem[Middelberg et al.(2008)]{middelberg08} Middelberg, E., Norris, R.~P., Cornwell, T.~J., et al.\ 2008, \aj, 135, 1276 
\bibitem[Murphy et al.(2013)]{murphy13} Murphy, T., Chatterjee, S., Kaplan, D.~L., et al.\ 2013, \pasa, 30, e006 
\bibitem[National Geographic (2005)]{natgeo}National Geographic Magazine, 2005, http://news.nationalgeographic.com/news/2005/04/0425\_050425\_hubble.html
\bibitem[Norris (1999)]{norris99}  Norris, R.P., 1999, Acta Astronautica, 47, 731.
\bibitem[Norris et al.(2006)]{norris06}Norris, R.P., et al. 2006, \aj, 132, 2409 
\bibitem[Norris et al.(2011)]{norris11} Norris, R.~P., Hopkins, A.~M., Afonso, J., et al.\ 2011, {\it EMU: Evolutionary Map of the Universe} \pasa, 28, 215 
\bibitem[Norris et al.(2013)]{Norris13}Norris, R.~P., Afonso, J., Bacon, D., et al.\ 2013, {\it Radio Continuum Surveys with Square Kilometre Array Pathfinders} \pasa, 30, 20  
\bibitem[Norris et al.(2015)]{norris15} Norris, R., Basu, K., Brown, M., et al.\ 2015, Advancing Astrophysics with the Square Kilometre Array (AASKA14), 86 
\bibitem[O'Brien et al.(2016)]{obrien} O'Brien, A.~N., Tothill, N.~F.~H., Norris, R.~P., \& Filipovi{\'c}, M.~D.\ 2016, arXiv:1602.01914 
\bibitem[Padovani et al.(2011)]{padovani11} Padovani P., Miller N., Kellermann K. I., Mainieri V., Rosati P., Tozzi P., 2011, \apj 740, 20
\bibitem[Park et al. (2017)]{park17} Park, L., Norris, R.P., \& Crawford, E., 2017, \pasa, in preparation.
\bibitem[Perlmutter et al.(1999)]{perlmutter99} Perlmutter, S., et al., 1999, \apj, 517, 565
\bibitem[Popper(1959)]{popper}Popper, K.,1959, The Logic of Scientific Discovery. New York, NY: Basic Books.
\bibitem[Rees et al.(2016)]{rees16} Rees, G.~A., Spitler, L.~R., Norris, R.~P., et al.\ 2016, \mnras, 455, 2731 
\bibitem[Rees et al.(2017)]{rees17} Rees, G., et al., 2017,  in preparation
\bibitem[Riess et al.(1998)]{riess98} Riess, A. G.,  et al., \aj, 116, 1009
\bibitem[Riggi et al.(2016)]{riggi16} Riggi, S., Ingallinera, A., Leto, P., et al.\ 2016, \mnras, 460, 1486 
\bibitem[Springel et al. (2005)]{springel05}Springel, V., et al., 2005, Nature, 435, 629
\bibitem[Sutherland \& Saunders(1992)]{sutherland92} Sutherland, W., \& Saunders, W.\ 1992, \mnras, 259, 413 
\bibitem[van der Kruit(1971)]{vdk71} van der Kruit, P.~C.\ 1971, \aap, 15, 110 
\bibitem[Weston et al.(2017)]{weston17} Weston, S., et al., 2017,  in preparation
\bibitem[Wilkinson et al.(2004)]{wilkinson04}Wilkinson, P.~N., et al., 2004, New Astr. Rev., 48, 1551 
\bibitem[Wilkinson(2007)]{wilkinson07} Wilkinson, P.\ 2007, From Planets to Dark Energy: the Modern Radio Universe, 144 
\bibitem[Wilkinson(2015)]{wilkinson15} Wilkinson, P.\ 2015, Advancing Astrophysics with the Square Kilometre Array (AASKA14), 65 
\bibitem[Williams et al.(1996)]{williams96} Williams, R.~E., Blacker, B., Dickinson, M., et al.\ 1996, \aj, 112, 1335 
\bibitem[Williams et al.(2000)]{williams00} Williams, R.~E., Baum, S., Bergeron, L.~E., et al.\ 2000, \aj, 120, 2735 
\bibitem[Woit(2011)]{woit11}Woit, P., 2011, Not Even Wrong: The Failure of String Theory and the Continuing Challenge to Unify the Laws of Physics. Random House.
\end{thebibliography}
\end{document}